\documentstyle[prl,aps,twocolumn,epsf]{revtex}

\draft

\begin{document}

\twocolumn[\hsize\textwidth\columnwidth\hsize\csname
@twocolumnfalse\endcsname

\title{Magnetic field induced localization\\
in a two-dimensional superconducting wire network}

\author{C.C. Abilio, P. Butaud, Th. Fournier and B. Pannetier}
\address{Centre de Recherches sur les Tr\`es Basses
Temp\'eratures-C.N.R.S. associ\'e \`a l'Universit\'e Joseph Fourier\\
25 Av. des Martyrs, 38042 Grenoble Cedex 9, France}
\author{J. Vidal}
\address{Groupe de Physique des Solides-CNRS, UMR 7588,
Universit\'{e}s  Paris 7 et Paris 6,\\
2 place Jussieu, 75251 Paris Cedex 05, France}
\author{S. Tedesco and B. Dalzotto}
\address{LETI (CEA-Grenoble), 17 Av. des Martyrs, 38054 Grenoble Cedex 
9, France}

\maketitle
\begin{abstract}
We report transport measurements on superconducting wire networks
which provide the first experimental evidence of a new localization 
phenomenon induced by magnetic field on a 2D periodic structure.
In the case of a superconducting wave function this phenomenon 
manifests itself as a depression of the network
critical current and of the superconducting transition temperature
at a half magnetic flux quantum per tile. In addition, the strong 
broadening of the resistive transition observed at this field 
is consistent with enhanced phase fluctuations due to this 
localization mechanism.
\end{abstract}

\pacs{PACS numbers: 72.15, 73.23, 74.25}

]

\bigskip

\narrowtext
%
%%%%%%%            INTRODUCTION
%
In a recent paper \cite{Vidal} a novel case of extreme localization 
induced by a transverse magnetic field was predicted for non 
interacting electrons in a two-dimensional (2D) periodic structure.  
This new phenomenon, due to a subtle interplay between lattice 
geometry and the magnetic field, differs from Anderson localization on 
two essential points: it occurs in a pure system, without disorder, 
and the system eigenstates are not localized but non-dispersive 
states.  In a tight-binding (TB) approach, it can be simply understood 
in terms of Aharonov-Bohm effect which, at half a flux quantum per 
unit tile (half-flux), leads to fully destructive quantum 
interferences.
% For this flux, the electrons motion is confined in the 
% so-called Aharonov-Bohm cages resulting from an effective cancellation 
% of some hopping terms. 
For this flux, the set of sites visited by an initially localized 
wave-packet will be bounded in Aharonov-Bohm cages \cite{Vidal}.
This effect is absent on other regular 
periodic lattices at half-flux, such as the square and the triangular 
lattices.

Superconducting wire networks are suitable to address phase 
interference phenomena driven by a magnetic field\cite{theo}.  These 
systems are extremely sensitive to phase coherence of the 
superconducting order parameter over the network sites which is 
exclusively determined by the competition between the external field 
and the network geometry\cite{exp}.  Besides, the quantum regime is 
accessible even in low $T_{c}$ diffusive superconductors: since all 
Cooper pairs condense in a quantum state, the relevant wavelength is 
associated with the macroscopic superfluid velocity and can be much 
larger than the lattice elementary cell\cite{2DEG}.  Also, the 
magnetic field corresponding to one superconducting flux quantum 
$\Phi_{o}=hc/2e$, is easily accessible: it is about $1~$mT for a 
network cell of $1~\mu$m$^{2}$, in contrast to the unattainable 
10$^{3}$~T for an atomic lattice.  In addition, some features of the 
TB spectrum, namely the Hofstadter butterfly\cite{Hofstadter}, are 
experimentally accessible in the model system of a superconducting 
wire network\cite{exp,honeycomb}.  As shown by de Gennes and 
Alexander\cite{deGennes,Alexander}, the linearized Ginzburg-Landau 
(GL) equations for a superconducting wire network can be mapped onto 
the eigenvalues equation of a TB hamiltonian for the same geometry.  
This mapping is of particular relevance since one of the remarkable
findings of Ref.\cite{Vidal} is the total absence of dispersion in the 
TB spectrum at half-flux.  In the context of a superconducting 
network, the localization effect is expressed by the inability of the 
superconducting wave function to carry phase information throughout 
the network and therefore, transport anomalies are expected.

\begin{figure}
		  \epsfxsize=8.0cm
		  \centerline{\epsffile{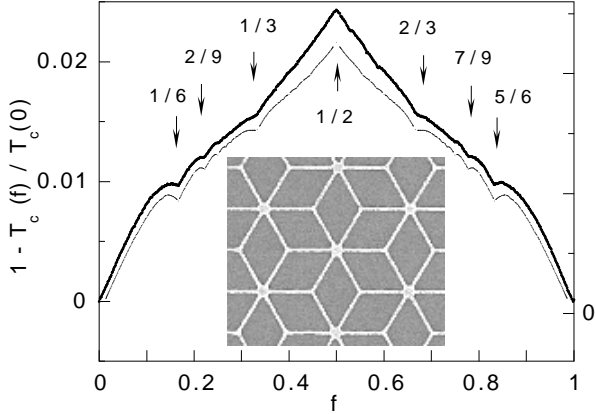}}
		  \caption{Reduced critical temperature \textit{vs} frustration
		  for Star~600 (heavy line, left axis).
		  The theoretical curve has been offset by 
		  -0.001 for clarity (small dots, right axis).
		  Inset: Electron micrography of the
		  $T_3$ network ($a=1.0~\mu$m). Measuring
		  current is applied along the horizontal direction.}
		  \label{Tc}
	  \end{figure}

In this Letter we present transport measurements on 2D 
superconducting networks with the so-called
$T_{3}$ geometry (see inset of 
Fig.~\ref{Tc}). Our results allow us to confirm some of
the exotic features of the $T_{3}$ energy spectrum related to
the localization mechanism. The field-temperature (H,T) superconducting
transition line is determined and related to the ground state
of the $T_{3}$ spectrum.
We also compare the critical current as a function of the magnetic 
field with calculations of the group velocity.
The striking behavior found at half-flux is discussed
as a possible signature of localization effects.
The strong broadening of the normal to
superconductor transition supports this interpretation. Very few 
experiments were reported so far on localization phenomena in 
superconducting networks and only the issues of irrational magnetic 
flux\cite{Chaikin} or disorder \cite{Goldman} have been addressed. 

%%%%%%%%%%%         SAMPLE DETAILS
%
%
The networks pattern was defined on a 600~nm thick layer of
positive UV3 resist using an \textit{e-beam} writer Leica VB6-HR.
A $100$~nm thick layer of pure aluminum was
\textit{e-beam} evaporated in a ultra-high vacuum chamber, followed
by the resist \textit{lift-off}\cite{Shipley}.
We designed two series with a large patterned area:
Star~600 defined on a  $0.6\times 1$~mm$^{2}$ surface and Star~20 defined 
on a surface of  $0.02\times 1$~mm$^{2}$,
which required stitching of $200\times 200~\mu$m$^2$ writing fields.
The elementary tile side length is $a=1~\mu$m, the wires having 100~nm width and 100~nm thickness.
%----------------     Fig 1 ------------
%-----------------------------------------
%
Dynamic resistance
measurements were performed using a 33~Hz ac four terminal resistance 
bridge with an ac measuring current of 20~nA. 
Sample probes were connected to the
cryostat terminals by ultrasonic bonding of $25~\mu$m gold wires.
Non invasive voltage probes were placed at 0.2~mm from the current 
pads.
The zero field transition temperatures $T_{c}(0)$
were 1.234~K for Star~600 and 1.240~K for Star~20, using
a resistance criteria of half the normal state resistance $R_{n}$ at 
1.25~K, which are $4.20~\Omega$ and $63.56~\Omega$, respectively.
The resistive transition width in zero field is 3~mK ($10\%-90\%$)
for both samples indicating 
a good homogeneity of the networks.

%
%
%%%%%%%%%%%%       CRITICAL TEMPERATURE TRANSITION 
%

The field dependent transition temperature $T_{c}$(H) was monitored  by locking
the temperature controller to keep the sample resistance at 
$0.5R_{n}$ as the magnetic field is varied.
The experimental data is to be compared with the lowest energy solution of the network linear 
GL equations, that is given in terms of the ground state eigenvalue 
$\epsilon_{g}(f)$ of 
the TB spectrum \cite{Alexander} by,
%
%------------    Equation 1
      \begin{equation}
		  1-\frac{T_{c}(f)}{T_{c}(0)}=\frac{\xi(0)^{2}}{a^{2}}
		  \arccos^{2}\left(\frac{\epsilon_{g}(f)}{\sqrt 18}\right),
		  \label{Tc-vs-energy}
	  \end{equation}
where $\xi(0)$ is the superconducting coherence length at zero 
temperature and $f$ the frustration.  Neglecting field screening 
effects, $f=\Phi/\Phi_{o}$ where $\Phi = Ha^{2}\sqrt{3}/2$ is the 
magnetic flux through a rhombus tile.

The transition line of Star~600 is plotted in Fig.~\ref{Tc} in reduced 
units $1-T_{c}(f)/T_{c}(0)$ as a function of frustration.  Since the 
transition line is periodic on $\Phi_{o}$ we only displayed it in the 
field range $0<f<1$\cite{exp}.  A small parabolic background due to 
field penetration in the wires was subtracted from the experimental 
$T_{c}(f)$.  We also display the theoretical $T_{c}(f)$ obtained using 
Eq.~(\ref{Tc-vs-energy}) and $\xi(0)=157$~nm, the only adjustable 
parameter.  The fine field structure of the experimental data is very 
well described by the theoretical curve.  Distinct downward cusps are 
visible at low order rationals $f=1/q$, for $q=3,4,6$, and 2/9.  They 
reflect the long range phase ordering of the order parameter among 
network sites, established at fields commensurate to the underlying 
lattice.  These features were discussed previously\cite{exp,Chaikin}.  
The novel feature of the transition line occurs at $f=1/2$, where the 
maximum of $T_{c}(f)$ depression (30~mK) is achieved, associated with 
an inversion of the field modulation concavity.  This anomalous cusp 
persists distinctly at all criteria used on $T_{c}(f)$ determination, 
from $0.06R_{n}$ to $0.87R_{n}$, though the downward cusps at other 
rationals fade out with increasing temperature.  This cusp is similar 
to the $T_{c}$ variation in a \underline{single} loop geometry close 
to $f=1/2$~\cite{LP}, and is characteristic of quantum effects 
determined on a finite length scale.  It indicates that at half-flux 
the network transition is determined by fluxoid quantization at 
independent tiles.

The maximum depression of $T_{c}(f)$ at half-flux shows the strong
incommensurability at this field.
To our knowledge, these results are the first experimental observation 
of such an effect on an extended periodic network.
Besides, they indicate that 2D periodicity is not a sufficient 
condition for a commensurate state to exist at rational $f$.

%
%
%%%%%%%%            TRANSITION WIDTH
%
%%%____________           Fig 2  transition  width
       \begin{figure}
		   \epsfxsize=8.0cm
		   \centerline{\epsffile{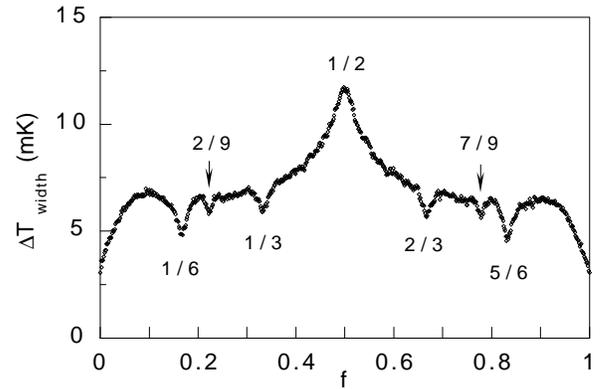}}
		   \caption{Transition width $vs$ frustration for Star~600.
		   The large broadening close to $f=1/2$ indicates the presence of
		   strong phase fluctuations.}
		   \label{transition-width}
	   \end{figure}

We also observed
a strong broadening of the resistive transition $\Delta 
T_{width}$,
at half-flux, as displayed in Fig.~\ref{transition-width} for Star~600.
$\Delta T_{width}$ is obtained as the difference between 
the $T_{c}(f)$ curves taken for criteria $0.6 R_{n}$ and $0.1 R_{n}$, respectively.
The anomalous enhancement (up to 12 mK) at half-flux,
twice the average width over most of the field range,
confirms the singular behavior found at this field.
At the strong commensurate fields
$f=0,\frac{1}{6},\frac{1}{3}$, $\Delta T_{width}$
is sharply reduced to a few mK as expected for 
a phase ordered system.
Close to these fields, the phase of the order parameter
at the network sites is able to "lock" in the nearest commensurate state with the
creation of few mobile defects, broadening slightly the transition.
Close to half-flux no commensurate state is available, thus phase 
correlations between network sites cannot be established, leading to a strong broadening of the transition.

In fact, the $T_{3}$ tiling geometry can be viewed as an
ensemble of three coupled triangular sublattices, two formed by the 3-fold 
sites and another by the 6-fold sites.
The singular properties of the $T_{3}$ spectrum
$\epsilon(f)$, at frustration $f$ are simply revealed by the transformation:
%
%_____________       Equation energy spectrum
      \begin{equation}
		  \epsilon^{2}(f)-6=2\cos({\pi f})~\epsilon_{T}(f_{T})
		  \label{star-tri}
	  \end{equation}
that relates $\epsilon(f)$ to the triangular lattice
eigenvalues $\epsilon_T(f_{T})$ at frustration $f_T=3f/2$\cite{Claro}.
At half-flux, due to cancellation of the cos(${\pi f}$)
prefactor, all the energy levels collapse into two highly degenerate 
discrete levels at $\epsilon=\pm \sqrt{6}$, forming flat, non-dispersive 
bands, in addition to the $\epsilon=0$ flat band.
%
%
%%%%%%%%%%%%%%%%%%    CRITICAL CURRENT
%
%
Due to the mapping of the TB problem onto the linearized GL approach,
the superfluid velocity can be expressed in terms of the group
velocity of the band spectrum close to the ground state.
In the context of a superconducting wave function, a non-dispersive 
state cannot carry phase information through the network, contrary to 
a Bloch state.  Therefore, critical current measurements give 
information on the network ability to sustain a supercurrent, {\it 
i.e.}, both a finite order parameter and a finite superfluid velocity.

The critical current was studied as a function of field from the 
dynamic resistance characteristics $vs$ increasing dc bias current at 
temperatures close to $T_{c}(0)$.  The used criteria was the threshold 
current for which the dynamic resistance exceeds $0.2\%~R_{n}$.  
Within the sensitivity limits of our measurements, it corresponds to the 
maximum current that the circuit is able to carry without dissipation.  
To avoid heating effects due to feeding a large current, we 
used sample Star~20 with 23 cells (20~$\mu$m) width.  The critical 
current density per wire $J_{c}(T,f)$ is obtained from the network 
critical current divided by the number of parallel wires (25) and the 
wire cross section.

Close to $T_{c}$, we expect the critical current to follow
a $3/2$ power law that generalizes the depairing current of a one-dimensional
superconducting wire \cite{Tinkham} to a superconducting 
network\cite{buisson}:
%
%________________    Equation critical current
\begin{equation}
J_{c}(T,f)=J_{n}C(f)\left(\frac{T_{c}(f)-T}{T_{c}(0)}\right)^{3/2},
\label{jc}
\end{equation}
where $J_{n}$ is the zero field depairing current density at T~=~0~K. 
The field dependent coefficient $C(f)$ is derived from the band 
curvature, $\partial^{2}\epsilon_{g}/\partial{k}^{2}$ close to the 
ground state, $\epsilon_{g}(f)$ by,
%
%_______________     Equation 
      \begin{equation}
		  C^{2}(f)=-\frac{1}{a^{2}\sqrt{18-\epsilon_{g}^{2}}}
		  \frac{\partial^{2}\epsilon_{g}}{\partial {k}^{2}}
		  \arccos{\frac{\epsilon_{g}}{\sqrt18}}
		  \label{courbure}
	  \end{equation}
%
% where
% $\partial^{2}\epsilon_{g}/\partial{k}^{2}$
% is the band curvature close to the ground state $\epsilon_{g}(f)$.

\begin{figure}
		  \epsfxsize=8.0cm
		  \centerline{\epsffile{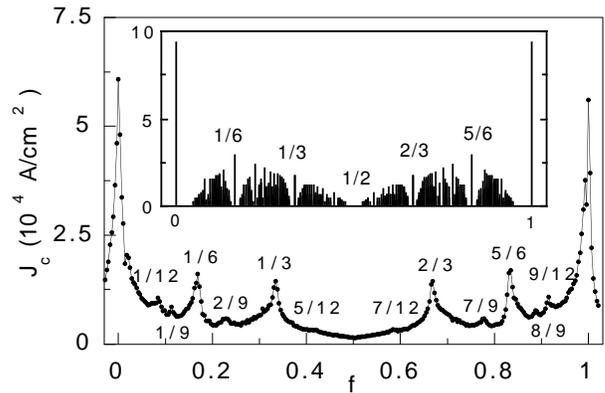}}
		  \caption{Critical current density of Star~20 as a function of
		  $f$ at T=1.185~K. The 
		  depression at half-flux is the signature of the 
		  non-dispersive state.
		  Inset: Theoretical values at the same temperature.}
		  \label{Ic}
	  \end{figure}

In Fig.~\ref{Ic} is displayed the field dependence of $J_{c}$, 
at $T=0.96~T_{c}(0)$ (1.185~K).
Sharp peaks are obtained for the same frustrations as the 
downward cusps observed in the transition line. The remarkable finding is the total 
absence of peak in the critical current at the lowest order rational $f=1/2$,
exhibiting a clear minimum at this field. For all studied temperatures 
the critical current was always found to exhibit the lowest values at $f=1/2$.
%
%___________________      Fig 3 
%__________________________________

In the inset of Fig.~\ref{Ic} is displayed the theoretical $J_{c}$ 
obtained using (\ref{jc}) and (\ref{courbure}) for $f=p/q$, $q<50$.  
We used $T=0.96~T_{c}(0)$ and $J_{n}=7 \times 10^6$~A~cm$^{-2}$, 
estimated from a 3/2 power law fit of the zero field $J_{c}(T,0)$ 
data.  The critical current consists of successive $\delta$-functions 
at rational frustration $f=p/q$ except at 1/2.  The highest values of 
$J_{c}$ were obtained for $f=0,\frac{1}{6},\frac{1}{3}$ and the 
symmetric values.  For these rational frustrations the spectrum is 
band-like and the group velocity is finite (Bloch states).  The same 
applies for other regular periodic lattices such as the square lattice 
where a checkerboard commensurate state leads to an important peak at 
f=1/2~\cite{buisson}.  However, in the 
$T_{3}$ case at half-flux the group velocity is strictly zero due to 
the absence of dispersive states and the critical current vanishes.  
This situation is original for an infinite tiling and is due to the 
special $T_{3}$ geometry.  A similar effect is found when $f$ 
approaches irrational frustration (for example, at small frustration 
$f=1/q$, with large $q$): the bandwidth then becomes exponentially 
small and therefore the group velocity and the critical current are 
suppressed.

The experimental data follow the same qualitative 
behavior {\it vs} field as the theoretical predictions.
The $3/2$ power dependence of $J_{c}$ {\it vs} temperature
at constant magnetic field was observed
in the temperature range $T_{c}(f)-T<20$~mK.
Namely, the field dependent coefficient $C(f)$ at half-flux
is reduced to $17\%$ of its zero field value.
The $C(f)$ depression, thus of $J_{c}$, reflects 
the effect of the band structure on the superfluid velocity, and
provides a strong evidence of the non-dispersive character of 
the state at $f=1/2$, although the measured critical
current does not vanish.

One possible explanation for the incomplete suppression of 
$J_{c}$ is the network finite size.
A current carrying state (an edge state 
similar to surface superconductivity in finite type-II superconductors) 
exists along each edge of the finite network and is expected to lead to 
a non-zero supercurrent. A second possible origin for finite $J_{c}$ is 
the influence of the GL non-linear term which was neglected in 
Eq.~(\ref{jc}). Presumably, the non-linear terms in the GL formulation are 
responsible for degrading the fine features of the band structure 
and therefore give a finite critical current. 
The critical current observed in Fig.~\ref{Ic} at small frustrations,
for example close to zero, may have the same origin. To go further, 
an exact solution of the non-linear GL equations would be needed. 
Nevertheless, as demonstrated by Abrikosov\cite{abrikosov}, a 
good physical insight of the superconducting properties can be 
obtained from the eigenstates of the linearized GL equation. 

%
%
%%%%%%%%%%%%%%       VORTEX LATTICE 
%
This phenomenon suggests interesting properties 
of the vortex sublattice. In this context, the coupling between network 
sites can be expressed as a landscape of energy barriers against vortex motion.
For example, at
$f=1/3$, a periodic vortex configuration can be easily constructed,  
matching perfectly the underlying lattice.
This configuration is strongly pinned and very stable against driving currents,
leading to a large critical current.
The decoupling of some network sites at $f=1/2$ suggests that,
in the absence of pinning,
the vortex configuration will be highly disordered.
Therefore, a significant dissipation is expected for small driving 
currents, as revealed on our experiments by the suppression of critical current 
and the anomalous transition broadening. These considerations 
are supported by preliminary experiments on vortex
decoration which indicate a highly disordered vortex distribution at $f=1/2$
and will be addressed elsewhere.

More subtle is the commensurate state at $f=1/6$, which corresponds to 
the $f_{T}=1/4$ state of the 
triangular lattice formed by the 6-fold sites (see Eq.~\ref{star-tri}).
As shown in Ref.~\cite{triangular}, the uniformly frustrated XY model on a triangular 
lattice at $f_{T}=1/4$ presents an accidental degeneracy of the 
ground state with zero energy domain walls which can weaken the 
global phase coherence. In our experiments we do 
observe a critical current peak at $f=1/6$, almost as large
as at $f=1/3$.
The singular behavior observed experimentally at $f=1/2$ ($f_{T}=3/4$)
is completely absent at $f=1/6$ ($f_{T}=1/4$) and therefore, cannot be 
simply related to the triangular lattice problem.
Besides, it is not clear if the accidental degeneracy 
persists for a tight binding coupling.

In summary, the anomalous transport behavior of the $T_{3}$ 
superconducting networks at half-flux is consistent with the 
localization effect predicted in Ref.~\cite{Vidal}.  The transition line is in excellent 
agreement with the related $T_{3}$ ground state.  The broad transition 
width at half-flux, indicates a strong enhancement of phase 
fluctuations which we assign to destructive quantum interferences at 
this field.  The reduction of the critical current at $f=1/2$, which 
was never observed so far in periodic superconducting networks, 
illustrates the inability of the network to sustain a transport 
current. This behavior is the analog, in the superconductor case, of 
the metal-insulator transition predicted in Ref.~\cite{Vidal}.

We acknowledge B. Dou\c{c}ot, R. Mosseri and O. Buisson for fruitful 
discussions.  The \textit{e-beam} lithography was carried out with 
PLATO organization teams and tools.
C.C.A. is supported by a grant from the Portuguese Ministry for 
Science and Technology.
Discussions within the TMR n$^o$FMRX-CT97-0143 are acknowledged.

%
%
%
%    References
%

\end{document}